\def\boxit#1{\vbox{\hrule\hbox{\vrule\kern3pt
\vbox{\kern3pt#1\kern3pt}\kern3pt\vrule}\hrule}}

\def\bea{\begin{eqnarray}}
\def\be{\begin{equation}}

\def\eg{{\it e.g.}}
\def\eea{\end{eqnarray}}
\def\ee{\end{equation}}
\def\eps{\varepsilon}
\def\epsp{\varepsilon^{\prime}}

\def\expectop#1#2#3{{ \langle {#1} | {#2} | {#3} \rangle }}

\def\fract#1#2{{#1}\,/{#2}}

\def\ie{{\it i.e.}}

\def\rarrow{\rightarrow}
\documentstyle[12pt,worldsci,epic,eepic,epsf]{article}
\begin{document}
\hfill UTPT-92-13
\title{{\bf K and B Physics: a way beyond the Standard Model }}
\author{Patrick J. O'Donnell
\thanks{Invited talk presented to the XV
International Warsaw meeting, Kazimierz, Poland, May25-29, 1992.}\\
{\em Department of Physics, University of Toronto, 60 St. George
Street, \\
Toronto, Ont. M5S 1A7, Canada}}

\maketitle
\setlength{\baselineskip}{2.6ex}

\begin{center}
\parbox{13.0cm}
{\begin{center} ABSTRACT \end{center}
{\small \hspace*{0.3cm}
I review some aspects of $K$ and $B$ physics both in
the context of the standard model and in some cases in a scenario
which is rather different from the standard model. I discuss, in
particular, where we are likely to see deviations from the
standard model in the near future before new colliders are
built.
}}
\end{center}

\section{The Three Family Standard Model.}

In the title, which was given to me by the organizers, I interpret
the word way as in path. However it also appears that the title
somehow implies that new physics means the standard model is wrong.
What if it is right?  What would be the path in this case? It is like
the story of a visitor to a remote part of Ireland who gets
lost. He comes to a cross road where he sees an old man sitting. After
passing the time of day with the old man he then asks for
directions to the village he wants to visit. The old man thinks for a
while and then remarks ``If I wanted to go to there, I would not
start from here."

\subsection{Unitarity Constraints.}

The standard model, with three families has the interaction
Lagrangian

\begin{equation}
{\cal L}_{I} = - \frac{g}{\sqrt{2}} (\bar{u}\hspace{0.2cm}
\bar{c} \hspace{0.2cm} \bar{t} \hspace{0.02cm})_{\rm L}
\gamma_{\mu}W^{\mu} V
\left( \begin{array}{c}
d                                  \\ [1ex]
s              \\ [1ex]
b                             \\
\end{array} \right)_{\rm L} + h.c.  \label{Li}
\end{equation}
where $g$ is related to the Fermi constant by
\begin{equation}
\fract{g^{2}}{8m_{W}^{2}} = \fract{G_{F}}{\sqrt{2}}
\end{equation}
Here $V$, the CKM\cite{C,KM} matrix is a unitary matrix connecting
the mass and weak eigenstates. It is important to note that it
contains information about the left hand quarks only. The right
hand sector is quite independent of it. For $n$ doublets of quarks the
CKM matrix is the smallest unitary matrix which can account for all
of the mixing among the families and at the same time for CP
violation. Remember, for a unitary matrix with $n$ rows and columns
the number of parameters is 1 for $n = 2$, the Cabibbo angle, and 4,
consisting of three angles and a complex phase for $n = 3$, the CKM
matrix. In
general there are $(n-1)^{2}$ parameters which divide up into
$n(n-1)/2$ real angles and $(n-1)(n-2)/2$ phases. Lack of
unitarity of the $3 \times 3$ matrix would be a sure sign of
something beyond the standard model. At the same time, extra
families obviously introduce a much larger number of parameters,
e.g., for $n = 4$ there would now be 9 parameters in total, with 6
real angles and 3 phases.

\subsection{The CKM Matrix and Model Dependencies.}

We have the CKM\cite{C,KM} matrix, in the Wolfenstein form\cite{Wolf1}
\begin{equation}
V=\left( \matrix{V_{ud} & V_{us} & V_{ub} \cr
V_{cd} & V_{cs} & V_{cb} \cr
V_{td} & V_{ts} & V_{tb} \cr} \right) \simeq \left(
\matrix{1-{\lambda^2 \over 2} & \lambda & A\lambda^3(\rho-i\eta) \cr
-\lambda & 1-{\lambda^2 \over2} & A\lambda^2 \cr
A\lambda^3(1-\rho - i\eta) & -A\lambda^2 & 1 \cr}\right) + O(\lambda^4)\\
\label{matrix}
\end{equation}

The present limits are\cite{PDG}
$\mid V_{us}\mid  \equiv  \lambda  \simeq  \sin \theta _{c} = 0.2205 \pm
0.0018$,
$\mid V_{cs}\mid  = 1.02 \pm  0.18$, $\mid V_{cd}\mid  = 0.204 \pm  0.017$
and $\mid V_{ud}\mid  = 0.9744 \pm  0.0010$.
There has been new activity in the determination of $\mid V_{cb}\mid $,
which now has the value\cite{B} $0.040 \pm  0.003 \pm  0.002$, or
equivalently, from the definition in Eq. \ref{matrix},
$A = 0.90 \pm  0.10$. A number of
attempts to deduce the value of $V_{cb}$ in a model independent way have
recently been given\cite{B,R,NR,N1,N2}. There has also been criticism of
some of these attempts\cite{RT}. A recent  phenomenological fit to the data
with a parametrization of the Isgur-Wise form factor $\xi (y)$ of the
form\cite{B} $\xi (y) = 1-\rho ^{2}(y-1)+c(y-1)^{2}$, finds
$\rho  = 1.10 \pm  0.10$ and $c = 0.67 \pm  0.25$.
Some of the earlier fits\cite{NR,N1,N2} violate the unitarity
constraints\cite{RT} on the slope of $\xi (y)$.

The elements of the CKM matrix, with $\eta$ denoting the amount of CP
violation lead to a (unitary) triangle in the standard model.
The size of the expected CP violation parameters $\eps,
\eps^{\prime}$ are given by the area of the triangle (in
units of $2A\lambda^{3}$). Thus, an accurate determination of a
non-zero value for $\eta$ will be a sure sign of CP violation in the
mixing matrix. Such an effect is usually called indirect CP
violation and the standard model has both direct and indirect modes
of CP violation, as I shall discuss below.
\begin{center}
\begin{picture}(200,100)(0,0)
\drawline(0,0)(200,0)
\drawline(0,0)(80,70)
\drawline(80,70)(200,0)
\put(80,80){\makebox(0,0){($\rho,\eta$)}}
\put(83,60){\makebox(0,0){$\alpha$}}
\put(0,-10){\makebox(0,0){(0,0)}}
\put(14,6){\makebox(0,0){$\gamma$}}
\put(200,-10){\makebox(0,0){(1,0)}}
\put(180,5){\makebox(0,0){$\beta$}}
\put(180,30){\makebox(0,0){$\fract{V_{td}} {\lambda V_{cb}}$}}
\put(5,30){\makebox(0,0){$\fract{V_{ub}^{*}}{\lambda V_{cb}}$}}
\end{picture}
\end{center}
\vspace*{0.5cm}
\begin{center}
{\small Fig. 1.  The unitary triangle in the standard model.}
\end{center}

The present limits on the allowed values of $\eta,\rho$ are still
dependent on the mass of the top quark, and until this is known, we
do not know even if the triangle has the angle $\gamma < \pi /2$ or
$\pi /2 \leq \gamma  < \pi$. It is
only possible to give a region of validity, depending on the
uncertainties in our knowledge of $\mid \fract{V_{ub}}{V_{cb}} \mid
= 0.09 \pm 0.04$, the uncertainty in the bag constant $B_{K} =
\fract{2}{3} \pm \fract{1}{6}$, the $B - \bar B$ mixing
parameter\cite{ET,CB} $x_{d} \equiv \fract{\Delta
M}{\Gamma} = 0.708 \pm 0.085 \pm 0.071$, and the values
of the $b$ quark mass and the decay constant
$f_{B}$. For example, in a recent review, Pich\cite{P} used the values $m_{b} =
4.6 \pm
0.1 GeV$ and $f_{B} = (1.7 \pm 0.4)f_{\pi}$.  More likely,
$m_{b} = 4.9 \pm 0.1 GeV$ since the constituent quark and current
quark masses are not that different for the heavy quarks.
A lower value of $f_{B} = (1.35 \pm 0.2)f_{\pi}$
would represent the consensus among quark models\cite{O'D}
and factorization models\cite{Ros1} and perhaps even the latest lattice
results\cite{Soni} (but see below). Many of the possibilities
are given in Ref. \citenum{BH}.
\section{CP Violation and Rare Decays of Neutral K Mesons.}
CP violation in the $K$ system has been with us since 1964, when the
effect was first seen\cite{CCFT} and a full description can be found
in books\cite{K}. Here I shall briefly describe the relevant
notation.

The mass eigenstates $K_{L}$ and $K_{S}$ can be written, in the Wu
-- Yang phase convention\cite{WY}, in terms of
the two CP eigenstates $K_{1}$ and $K_{2}$, corresponding to $CP=+1$
and $CP=-1$, respectively, as
\begin{eqnarray}
K_{L}& =&\frac{ K_{2}+\eps K_{1}}{\sqrt{1+\eps^{2}}} \nonumber \\
K_{S}& =&\frac{ K_{1}+\eps K_{2}}{\sqrt{1+\eps^{2}}}
\label{KLKS}
\end{eqnarray}
where, in a CP -- invariant world, the CP eigenvalues are
\begin{equation}
K_{1} =\frac{ K^{\circ} - \bar{K}^{\circ}}{\sqrt{2}} \; \hspace{1 cm}
K_{2} =\frac{ K^{\circ} + \bar{K}^{\circ}}{\sqrt{2}}
\end{equation}
\subsection{Direct and Indirect CP Violation.}

The parameter $\eps$ can be related to the quantities $\eta_{+-}$
and $\eta_{\circ\circ}$ which are defined in terms of the ratio of the
amplitudes for
decays into CP -- forbidden and CP -- allowed $2\pi$ combinations as
follows:-
\begin{eqnarray}
\eta_{+-}&=&\fract{\langle\pi^{+}\pi^{-}\mid T \mid
K_{L}\rangle}{\langle\pi^{+}\pi^{-}\mid T \mid K_{S}\rangle} \nonumber \\
\eta_{\circ\circ}&=&\fract{\langle\pi^{\circ}\pi^{\circ}\mid T \mid
K_{L}\rangle}{\langle\pi^{\circ}\pi^{\circ}\mid T \mid K_{S}\rangle}
\end{eqnarray}

When all the complications of phases are fixed up (see
Rosner\cite{Ros2} for a very complete discussion on this) the
quantities $\eta_{+-}$ and $\eta_{\circ\circ}$ can be written in terms of
two small quantities $\eps$ and $\eps^{\prime}$:-
\begin{equation}
\eta_{+-}=\eps + \eps^{\prime} \; \hspace{1 cm}
\eta_{\circ\circ}=\eps -2 \eps^{\prime} \hspace{3 cm}
\end{equation}

If the world were CP invariant then both types of $\eta$ quantities
would be zero. The parameter $\eps$ is a measure of the CP
violation coming from the mixing in the mass matrix and is
called {\em indirect} CP violation. The parameter
$\eps^{\prime}$ give a measure of the CP violation in the decay
amplitudes, which is called {\em direct} CP violation. A non -- zero
value for this leads to the relation
\begin{equation}
\mid \fract{\eta_{\circ\circ}}{\eta_{+-}}\mid^{2}\approx 1-6
Re\left(\fract{\eps^{\prime}}{\eps}\right)
\end{equation}

At the moment there are two conflicting values for
$(\eps^{\prime}/\eps)$ with the CERN experiment\cite{Barr}, NA31,
reporting $(2.3\pm 0.7)\times 10^{-3}$ and the Fermilab experiment\cite{Win},
E731 finding a null result $(0.60\pm 0.69) \times 10^{-3}$. The first of
these would
indicate that CP violation takes place in both ways, through the
mass matrix and in the decay amplitudes, while the second has only
the indirect, mass matrix source. This latter case could be explained
in a number of ways, including the superweak model\cite{Wolf2}, which was set
up
to give CP violation only through the mass mixing.

The phases of $\eps$ and $\epsp$ are determined by the final state
interactions in the $\pi \pi$ system and are nearly equal to one
another\cite{Ros2} and to $\pi / 4$. In the standard model, $\eps$ is
calculated using box diagrams\cite{CL} and $\epsp$ comes from
penguin diagrams. The $\Delta S = 2$ Lagrangian from the box diagrams
give an effective $4$ -- quark operator which is usually calculated
in the vacuum insertion approximation,
\be
\expectop {\bar K^{\circ}}{(\bar s d)_{V-A}(\bar s
d)_{V-A}}{K^{\circ}}=(\fract 8 3) B_{K} {f_{K}}^{2}{m_{K}}^{2}
\ee
Here $B_{K}$ is the bag constant, which parameterizes the vacuum
approximation and which is the subject of considerable uncertainty.
In the discussion above on the CKM parameters I used the central
value of 2/3. The range of values in the literature range from $2/3 \pm 0.1$
in $1/N$ calculations\cite{1/N} through a much smaller value in hadronic
sum rule calculations\cite{SR}, slightly higher values in QCD sum
rules\cite{QCDSR} of $0.54 \pm 0.22$ and $0.87 \pm 0.20$ in a lattice
calculation\cite{LAT}. On the other hand, the decay constant is well
known\cite{PDG}, $f_{K}=159.8  \pm 1.4 MeV$. In the standard model
the penguin and box diagrams differ slightly in phase and it is
expected that $\epsp / \eps \approx 0.5 - 1 \times 10^{-2}$, \ie,
there should be some direct CP violation.

\subsection{Possible Origin of CP Violation.}

A number of new physics models have been discussed in the
literature and I do not want to repeat that here\cite{Ros2}. Instead,
I will discuss a radical new proposal\cite{CR} on the origin of CP violation.
This will indeed be beyond the standard model if it is proven
viable. There is a very different signal from the standard model --
and even most non--standard models.

In a recent paper\cite{CR} Chardin and Rax revive an old idea that
was dismissed\cite{Mor,Good} {\em before} the discovery of CP violation.
In an early article by Morrison\cite{Mor} it was pointed out that
the anti--gravity theory violated CPT invariance, the E\"{o}tv\"{o}s
experiment,
and energy conservation -- not a good start for any theory!
A little later, Good\cite{Good} discussed the same
problem  in the context of $K$ physics. He  showed that the
$K_{L}$--$K_{S}$ mass difference is a very sensitive measure of any
new physics that would cause a difference in the mass, weight or
potential energy of the system. In particular, since $\delta M_{K}
= 3.522 \times 10^{-6} eV$ and the potential energy of a $K$
at the earth is about $0.4 eV$ this imposes very stringent limits on
the change of the mass matrix. For example, if the neutral $K$ mass
matrix is written\cite{Bell} as:-
\[ \left(
\begin{array}{cc}
M_{K}+V(1+\delta ) & iW \\
iW & M_{K}+V(1-\delta )
\end{array}  \right) \]
where $V$ is the gravitational potential and $\delta$ represents the
fraction which changes sign in anti--gravity, then $\delta$ has to be
extremely small since $\eps=\fract{V\delta}{\Delta M}\approx
\gamma^{p}$. Here $\gamma$ is the usual Lorentz factor and $p$
depends on the type of field causing the violation. This
can be an important effect since experiments have been done up to a
value of $\gamma=100$. The values of $p$ are $0, 1, 2$ for scalar,
vector and tensor fields, respectively. Thus a vector field will have
a limit for
$\delta$ relative to a scalar field that is down by $\gamma$ while
a tensor field is
down by $\gamma^{2}$.

Despite these pessimistic arguments, the idea has
recently been discussed\cite{GN} and revived\cite{CR}. The latter
argument can be stated by using the dimensions of the system,
the size of the $K$ and the time needed
for mixing via weak interactions $\delta \tau \approx \fract \hbar {\Delta
M c^{2}}$. If the quark anti--quark
separation in a time $t\approx \delta \tau$ is larger than the size
of the $K$ then a large $K_{S}$ component could be regenerated. An
estimate of this can be given: $gt^{2}\approx \eps \times \fract
\hbar {M_{K}c}$ which gives $t\approx 1.7\Delta \tau$. This suggests
that the CP--violating parameter could be written as
$\eps\approx\fract{{g \Delta \tau}^{2} \hbar}{M_{K}c}$. Any type of
long range mechanism for CP--violation will have $\epsp = 0$ but
this one has a distinctive difference from, say, the superweak
model, in that the similar parameter for CP--violation in B mesons
will scale down by $10^{-3}$ from the values of $\eps_{K}$. Thus,
indirect CP--violation in $K$ mesons and a detection of CP in B
physics will certainly tell whether there is anything to this
proposal.

\subsection{Rare K Decays: Present and Future.}

Here, I shall review, briefly, the possibilities of using the rare decays of
$K$ mesons to either get bounds on some of the quantities discussed
above or to see something new. Most of the experimental limits are
still well above the expected standard model predictions.

$\bullet K^{+} \rarrow \pi^{+} \nu \bar \nu $: the present
limit\cite{PD} from BNL--E787 at $5 \times 10^{-9}$ is now about an
order of magnitude
greater than the expected rate. This mode would measure both $\rho$
and $\eta$. Until it is seen at the expected rate, it can only signal
new physics! However, the chances of this happening are now quite
slim -- when I gave this talk the limit was about an order of
magnitude larger.

$\bullet {K^{\circ}}_{L} \rarrow  \mu \bar \mu$: the measured
branching fraction of $(7.3 \pm 0.4) \times 10^{-9}$ seems to be
understood\cite{KRY} in terms of the process
$K_{L} \rarrow \gamma \gamma \rarrow \mu \bar \mu$.

$\bullet {K^{\circ}}_{L} \rarrow \pi^{\circ} \nu \bar \nu$: the
expected branching fraction is $< 10^{-11}$ which is well below an
old experiment which has been retroactively analyzed\cite{PDG} to
give a branching ratio limit of $7.6 \times 10^{-3}$.

$\bullet {K^{\circ}}_{L,S} \rarrow  \gamma  \gamma$: the expected\cite{HK}
ratio of $|\fract \epsp \eps | \approx 0.02$.

$\bullet {K^{\circ}}_{L} \rarrow \pi^{\circ} e^{+} e^{-}$: the present limit
is $5.5 \times 10^{-9}$. The theoretical calculations of this mode have
been controversial. Decays via the CP--violating mode
are expected to dominate and give a limit of about $10^{-11}$ or even
an order of magnitude smaller. There is some doubt about whether the
direct and non direct CP--violation parts are of the same order of
magnitude (in which case there could be significant interference).
A further controversy revolves around the $2 \gamma$ CP--conserving
intermediate state $K_{L}\rarrow \pi^{\circ}\gamma \gamma \rarrow
\pi^{\circ}e^+ e^-$. Here, the estimates are all over the map from the
largest\cite{MI} $10^{-10}$ to\cite{Seh} $10^{-11}$, to\cite{FR}
$10^{-12}$, to\cite{Don,EPD} $10^{-13}\sim 10^{-15}$.
In fact from a recent NA31 result\cite{PD} we now have $Br(K_{L}\rarrow\pi^{o}
\gamma \gamma) = (1.7 \pm 0.2 \pm 0.2) \times 10^{-6}$ so that
${K^{\circ}}_{L,S} \rarrow  \pi \gamma  \gamma \rarrow \pi^{\circ}e^+ e^- \le
4.5
\times 10^{-13}$ ;
the CP--violation mode is in fact the dominant one.

For future prospects in $K$ physics we shall have to wait for a new
facility such as the proposed KAON factory in Canada. An indication
of the type of mass limits possible in such a facility has been
recently given\cite{CK}.
{}From the LEP
talk at this meeting we see that the limits on the Higgs Scalar
already exceeds any of the limits expected at KAON.
\section{B Physics}

Although K meson physics has been an important source of information
leading to the formulation of the standard model, B meson physics
is much newer and full of promise. This can be seen in the relevant
numbers of papers with the letter K (1064) or B (1762) in their title as
listed in the SLAC database (by July 17, 1992). (Not all of these
deal with the mesons - some refer to algebras!).

\subsection{$f_{B}$ and CP Violation Asymmetries.}

One important constant which is not known at the moment is $f_{B}$.
There are a number of attempts to calculate this;
if we write $f_{B}=nf_{\pi}$ then the question at the moment is
whether $n\approx 1$ or $2$?  In the table we summarize the results
from different types of calculations.
\begin{center}
\bf Table 1. \rm

This summarizes the present state of a number of calculations
of the ratios of the decay constants in units of $f_{\pi}$; in some
references only a few of the decay constants are available. For
comparison, I also show the appropriate constants in $D$ decays.
\end{center}

\begin{center}
\begin{tabular}{|llllrl|}
\hline
$f_{B}$ & $f_{B_{\rm s}}$ & $f_{D}$ & $f_{D_{\rm s}}$ & Reference & Method \\
\hline
1.1 & 1.4 & 1.8 &   &\citenum{O'D} & hyperfine int.\\
\hline
0.8 & 1.2 & 1.33 & 1.8  &\citenum{BDHS} & lattice \\
0.9 & 1.1 & 1.4 & 1.6  & \citenum{LAT} & \\
&&1.4&1.7 &\citenum{DeGL} & \\
\hline
0.9 & 1.3 & 1.1 & 1.6  &\citenum{Kras} & potential model \\
0.6 & .6 & 0.9 & 1.0  &\citenum{Suz} & \\
1.7 & 1.9 & 1.4 & 1.5 &\citenum{CCCN} & \\
1.2 & 1.6 & 1.8 & 2.2 &\citenum{CG} & \\
\hline
1.4 & &1.7 & 2.0 &\citenum{DP} & sum rules \\
1.4 & 1.5 & 1.3 & 1.6  &\citenum{Nar} & \\
&&& & \citenum{Rei} & \\
0.9 & & 1.2& 1.5 & \citenum{Shif} & \\
\hline
1.5 &&1.7&& \citenum{MT} & rel. quark model  \\
\hline
1.1 & 1.4 & 1.7 & 2.1 &\citenum{Ros1} & factorization \\
\hline
\end{tabular}
\end{center}

To get some of these values, I
used $f_{D_{s}} = 276 $ from the estimated result\cite{BStone}
$f_{D_{s}} = 276 \pm 45 \pm 44$ since in
some cases only the ratios of the coupling constants were calculated.
The lattice results in the table are now a few years old. More
recent lattice calculations tend to get larger values for $f_{B}/f_{\pi}$,
viz., $f_{B}/f_{\pi}= 2.3$ (ref. \citenum{ALL}), $1.4 <
\fract{f_{B}}{f_{\pi}} < 1.9$ (ref. \citenum{Som}), and $f_{B}/f_{\pi} =
1.5$ (ref.\citenum{BS}). However, there are large errors and \eg,
in the latter case the quoted result is $f_{B} = 195 \pm10 \pm30 -
60$ where the first two are roughly the statistical and systematic
errors and the last error only has a minus sign, representing the
difference in two methods used\cite{BS}.

It is important for CP--violation (for recent reviews see, \eg \
\citenum{Ros2},
\citenum{PT} and \citenum{NQ}) that we get a good estimate on
$f_{B}$. For example, the important mixing ratios, $x_d$ and $x_s$
are significantly affected; in the standard model they arise from
the box diagrams dominated by $t$ quark exchange;
\be
x_d=\frac{\Delta M}{\Gamma}=\frac{G_{F}^{2}}{6\pi^2}\mid\!V_{td}\!\mid
^{2}M_{W}^{2}m_{b}B_{B}f_{B}^{2}\eta_{B}\tau_{B}E(x_t)
\ee
with a similar equation for $x_s$. Here, E(x) is a known
function\cite{IL,COD} shown below in Section 3.3, $\eta_{B}$ is a QCD
correction, $B_{B}$ is the appropriate bag constant and $\tau_{B}$
is the lifetime. As shown below, the mixing ratios also enter into
the CP--violating asymmetries.

\subsection{An Ambiguity in the Standard Model.}

In this section we consider the CP--violating asymmetries in some
detail, since it has recently been shown that there could be an
important ambiguity in the standard model.

The asymmetry in the decay $B\rarrow \psi K_{S}$ is
defined as
\be
A(\psi K_{S})=\frac{\Gamma (B^{\circ}\rarrow \psi K_{S})-
\Gamma({\bar B}^{\circ}\rarrow \psi K_{S})}
{\Gamma(B^{\circ}\rarrow \psi K_{S})+\Gamma({\bar B}^{\circ}\rarrow \psi
K_{S})}
\ee

In an actual experiment, the semi-leptonic decays tag the
identity of the particle-antiparticle, since $B^{\circ}\rarrow l^{-}+...$
while $\bar B^{\circ}\rarrow l^{+}+...$. When time ordered, there
are four possible combinations\cite{HS}

\begin{center}
\begin{tabular}{lllll}
$B^{\circ},{\bar B}^{\circ}$
& $\rarrow l^{-}(t_{1})$ & $\psi K^{\circ}_{S}(t_{2})$ & $t_{1} > t_{2}$ & (1)
\\
& $\rarrow l^{+}(t_{1})$ & $\psi K^{\circ}_{S}(t_{2})$ & $t_{1} > t_{2}$ & (2)
\\
& $\rarrow l^{-}(t_{2})$ & $\psi K^{\circ}_{S}(t_{1})$ & $t_{2} > t_{1}$ & (3)
\\
& $\rarrow l^{+}(t_{2})$ & $\psi K^{\circ}_{S}(t_{1})$ & $t_{2} > t_{1}$ & (4)
\\
\end{tabular}
\end{center}

The asymmetry $A(\psi K_{S})$ is then defined by:
\be
A(\psi K_{S})= \frac{(1) -(2)+(3)-(4)}{(1)+(2)+(3)+(4)}
\ee

In the standard model, the angles $\beta$ and $\alpha$  of Fig. 1 are
related to the asymmetries of the decays of the $B$ into final states
$\psi K_{S}$ and $\pi^{+} \pi^{-}$, respectively in the following
way\cite{DR,DDGN}:-
\bea
sin(2\beta) &= & - \left(\frac{1+x^{2}}{x} \right) A(\psi K_{S})\nonumber \\
sin(2\alpha) &= & - \left(\frac{1+x^{2}}{x} \right) A(\pi^{+} \pi^{-})
\eea
where the mixing fraction is $x=\fract {\Delta M}{\Gamma}=0.708 \pm0.085 \pm
0.071$.

In the superweak model, there is no direct CP--violating effect,
$\epsp=0$, nevertheless, $B-\bar B$ mixing can give rise to a
non-zero asymmetry.
Since $CP(\pi^{+} \pi^{-})=1$ and $CP(\psi K_{S})=-1$ the superweak
model would predict a change in sign between the two final
states\cite{GerN,BW}. That is, {\em if} $\alpha=\beta$, then the standard
model would have the same (negative) asymmetries in both cases whereas the
asymmetries have opposite signs in the superweak model.

It has been recently pointed out\cite{BW} that there could be a
situation in which the standard model is indistinguishable from the
superweak model (at least with regard to these asymmetries). This
occurs when $sin (2\alpha) + sin (2 \beta) = 0$. This means that the
standard model would mimic the superweak model for any $\rho > 0$
and with the constraint $\eta=(1-\rho)\sqrt{\fract{ \rho} {(2-\rho)}}$
\subsection{Rare $B$ processes - mixing}

Here, we consider $B-\bar B$ mixing with the spectator quark being a
$d$ quark. We have
\be
\frac {\Delta M}{\Gamma_{12}}=\left \{ \frac{[0.85\pm 0.05]}{1.1}\frac{4
E(x_t)}{3\pi x_b}\right \} \simeq 60.
\ee
where $x_q=\fract {m^2_{q}}{m^2_{W}}$ , the two numerical factors
representing QCD corrections\cite{BG,BSS} and
\bea
E(x)&=&x\left \{\frac{1}{4} +\frac{9}{4(1-x)}-\frac{3}{2(x-1)^2} \right
\}+\frac{3}{2}\left ( \frac{x}{(x-1)}\right )^3 ln(x)\nonumber \\
&\approx& 0.159+0.582x-0.015x^2
\label{eqB}
\eea
with the last approximation to the first line coming from a recent
determination\cite{BG}, and which agrees very well with the full
expression for $1\leq x_t\leq 9.5$. Notice, that as the limits on the
top quark mass have risen in the past few years, care should be taken in
using such approximations. For
example, just a few years earlier, another approximation\cite{AB}
was given, $E(x)\approx 0.75 x^{7/4}$. This widely
disagrees with the full form of $E(x)$ in the present case for which
$x_t \geq 1$.

The quantity $\fract {\Delta M}{\Gamma_{12}}$ is a good one for
showing something beyond the standard model.
The calculation of $\Gamma_{12}$ is well understood.  It comes from
cutting the box-diagrams, so it is tree--level W mediated and there
does not seem to be a competing model available. If the ratio in
Eq. (\ref{eqB}) is to be made smaller then something must happen to
the calculation of $\Delta M$.  Since we know that $\fract{\Delta M}{\Gamma}
\sim 0.7$ it is not a simple matter in the standard (or nearly
standard) model to make much of a change.
\section{Inclusive Rare Decays of B Mesons.}

Inclusive rare decays of the B meson, considered as decays of the
form $b\rarrow s \gamma$ and $ b \rarrow s g$ have been the subject
of a great deal of interest over the past few years\cite{bdecays}
In the following section, we look at the standard model calculation
and estimate the reliability of the calculations.
\subsection{The Standard Model.}

The original calculation\cite{COD}, based on the similar calculation\cite{IL}
in $d\rarrow s \gamma$, suggested that if the top quark was
light with $20 GeV \le m_t \le 60 GeV$, say, then the process
$b\rarrow s \gamma$ would be an excellent top quark ``mass meter''. It
was later pointed
out\cite{BBM,DLTES}, using a modified version of the work of Shifman et
al.,\cite{SVZKS} that the short distance QCD enhancement of the
$\sigma . F$ operator would cause an increase in the rate for
$b \rarrow s \gamma$. However, both groups ignored terms coming
from graphs with the photon attached to the down quark line and
also ignored mixing effects from graphs with an internal gluon. More exact
calculations came soon after\cite{GSW,GOSN1,CCRV,Mis,ChoG}.

We now have confidence that the QCD scaling of the coefficient functions
of the effective operators is understood and so we are able to estimate well
a number of quantities which are of interest
experimentally.  The most basic of these is the rare inclusive decay
$b\rightarrow
s\gamma$.  When the quark and photon fields are on shell, the only operator
which contributes is the magnetic-moment operator ${\cal O}_{5}$, in
the notation of Ref. \citenum{GOSN1}, whose
coefficient function is $\tilde{C}_{5}(m_{b})$.

The width for the free quark decay $b\rightarrow s\gamma$ is given by
\begin{equation}
\Gamma(b\rightarrow s\gamma) =
\frac{\alpha G_{F}^{2} m_{b}^{5}}{288 \pi^{4}}
\mid\!V_{ts}V_{tb}\!\mid ^{2} (\tilde{C}_{5}(m_{b}))^{2}
\end{equation}
where $\alpha = 1/137$.  The
dependence on $m_{b}$ and the CKM elements may be removed by
the usual trick of normalizing to b-quark semileptonic decay $b\rightarrow
ce\overline{\nu}_{e}$.  The width for this process is given by
\begin{equation}
\Gamma_{SL} =
\frac{G_{F}^{2}m_{b}^{5}}{192\pi^{3}}\mid\!V_{cb}\!\mid ^{2}
\rho \!\left( m_{c}^{2}/m_{b}^{2} \right)
\chi \!\left( m_{c}^{2}/m_{b}^{2} \right)
\end{equation}
where the phase-space factor $\rho$ depends on the non negligible ratio
$y=m_{c}^{2}/m_{b}^{2}$:
\begin{equation}
\rho(y) = 1 - 8y + 8y^{3} - y^{4} -12y^{2}\ln y
\end{equation}
and equals 0.447 for $m_{c}=$ 1.5 GeV and $m_{b}=$ 4.5 GeV.  The factor
$\chi$ is the one-loop QCD correction to the semileptonic decay
\cite{CMCPTF}:
\begin{equation}
\chi(y) = 1 - \frac{2\alpha_{s}(m_{b})}{3\pi} f(y) ,
\end{equation}
with
$f(m_{c}^{2}/m_{b}^{2}) \approx$ 2.4; this correction gives a modest 12 per
cent suppression to the semileptonic width.  Experimentally, the ratio
$\mid\!V_{ts}V_{tb}/V_{cb}\!\mid\approx$ 1, so the KM elements cancel.

The result for $b\rightarrow s\gamma$ normalized to semileptonic decay is
\begin{equation}
\frac{\Gamma}
{\Gamma_{SL}}=
\frac{2\alpha}{3\pi\rho\chi}(\tilde{C}_{5}(m_{b}))^{2}
\end{equation}

The branching ratio for $b\rightarrow s\gamma$ is obtained by multiplying
this by the semileptonic branching ratio, which is found experimentally to
be about 10 per cent.
In Fig. 2 the upper line gives the branching ratio evaluated at $\mu =
m_{b}$, and the lower line at $\mu = M_{W}$.  One sees that
QCD scaling from $M_{W}$ down to $m_{b}$ enhances the branching ratio.  The
magnitude of the enhancement depends on the top quark mass, and is smaller
at the larger values of $m_{t}$.  This is because
the GIM suppression is weaker at large $m_{t}$.  The GIM suppression is
partially ``undone'' by QCD, and because there is less of a
suppression to ``undo'' at large $m_{t}$, the size of the QCD effect
is smaller.

With the top quark mass expected to be somewhere
around 140 GeV, and including the QCD corrections,
means that the sensitivity to $m_{t}$ is reduced compared to the
original expectations\cite{COD}.
The decay becomes less useful as a top quark ``mass meter''.  On the other
hand this lack of sensitivity to an unknown parameter of the
standard model makes the decay a good place to look for signals of
new physics beyond the standard model. \\
\hfill
\epsffile[-60 0 220 198]{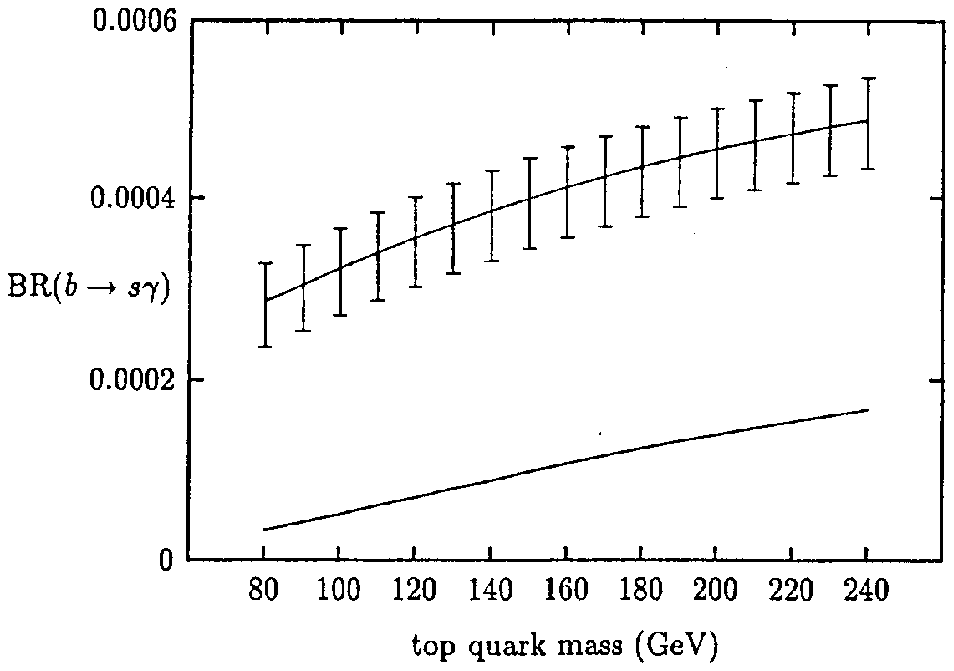}
\hfill
\begin{center}
{\small Fig. 2. The effects of QCD corrections.  }
\end{center}

The error bars in Fig. 2
reflect the fact that the QCD scale parameter
$\Lambda_{QCD}$ is not well known; the curve is plotted for a
central value of 200 MeV, with a range of $\pm$ 100 MeV.  Larger values of
$\Lambda_{QCD}$ correspond to larger values of the branching ratio.

The process $b\rightarrow s\gamma$ also receives a contribution
from\cite{GOSN1}
$b\rightarrow sg$. For all practical purposes, the effect is negligible.
Recently a corrected calculation for $b\rarrow sg$ was
given by Misiak\cite{Mis}. Also, a correction taking into
account the fact that the top quark is now thought to be much
heavier than the $W$ meson shows that the branching ratio increases
by as much as\cite{ChoG} $14\%$ at the higher values for $m_{t}$,
though still within the errors coming from $\Lambda_{QCD}$ shown in Fig.
2.

A simple numerical representation of the results of the QCD corrected inclusive
decay (central value in Fig. 2) is
\be
BR(b\rarrow s \gamma) = (2.314 + 0.5814 x_{t} - 0.033{x_{t}}^{2}) \times
10^{-4}
\ee
where, as usual, $x_{t}=\fract{m^2_{t}}{m^2_{W}}$. The most up-to
date limit on the branching ratio for the inclusive decay\cite{Dan} is $8.4
\times 10^{-4}$

\subsection{Other Models.}

I have not enough space to give more than a brief mention of
models beyond the standard model since, in many cases, there are a
number of parameters that can be adjusted. The major
extensions to the standard model are;

$\bullet$ Extra Families. With extra families the parameters can be
tuned to give rare $b$ decays with a branching ratio near to the
present bounds. The limit on the
number of neutrino families from the LEP data which now requires
that a fourth generation has $m_{\nu_4} > \fract {M_Z} {2}$ creates a
serious difficulty for these models.

$\bullet$ Supersymmetry. See the talk by Borzumati\cite{BG,BBM}.

$\bullet$ Extra Higgs, models with non tree-level Flavor Changing
Neutral Currents.
After $\eps$, $\fract \epsp \eps$ and mixing constraints are applied,
these keep the expected magnitude of the rare $b$ decays at the
order of the standard model\cite{GG}.

$\bullet$ Left--Right symmetry. These models need fine tuning. Like
SUSY, there are many variations.

\section{Exclusive Rare Decays in the B System}

There are limits on the exclusive decays of the $B$ meson which come
from four groups, ARGUS and Crystal Ball at DESY, CLEO at CESR and UA1
at CERN.  At the time
of writing, the best limits on the various decays were those listed in
the following table:

\begin{center}
\begin{tabular}{c|c|c|c}
Process & Limit & Experiment & Reference  \\ [1ex]
\hline \\ [-2ex]
$B\rightarrow K^{*}(892)\gamma$ & $2.4\times 10^{-4}$ &
	ARGUS & \citenum{ARG1} \\ [1ex]
$B^{\circ}\rightarrow K^{*}(892)^{\circ}\gamma$ & $0.92\times 10^{-4}$ &
	CLEO & \citenum{Dan} \\ [1ex]
$B^{+}\rightarrow K^{*}(892)^{+}\gamma$ & $5.2\times 10^{-4}$ &
	ARGUS & \citenum{ARG2} \\ [1ex]
$B^{+}\rightarrow K^{*}(892)^{+}\gamma$ & $3.7\times 10^{-4}$ &
	CLEO & \citenum{Dan} \\ [1ex]
$B^{\circ}\rightarrow K^{*}(1430)^{\circ}\gamma$ & $4.4\times 10^{-4}$ &
	ARGUS & \citenum{ARG2} \\ [1ex]
$B^{+}\rightarrow K^{*}(1430)^{+}\gamma$ & $1.3\times 10^{-3}$ &
	ARGUS & \citenum{ARG2} \\ [1ex]
\hline \\ [-2ex]
$B^{\circ}\rightarrow K^{\circ}e^{+}e^{-}$ & $1.5\times 10^{-4}$ &
	ARGUS & \citenum{ARG3} \\ [1ex]
$B^{+}\rightarrow K^{+}e^{+}e^{-}$ & $5\times 10^{-5}$ &
	CLEO & \citenum{CLEO2} \\ [1ex]
$B^{\circ}\rightarrow K^{\circ}\mu^{+}\mu^{-}$ & $2.6\times 10^{-4}$ &
	ARGUS & \citenum{ARG3} \\ [1ex]
$B^{+}\rightarrow K^{+}\mu^{+}\mu^{-}$ & $1.5\times 10^{-4}$ &
	CLEO & \citenum{CLEO2} \\ [1ex]
$B^{\circ}\rightarrow K^{*}(892)^{\circ}e^{+}e^{-}$ & $2.9\times 10^{-4}$ &
	ARGUS & \citenum{ARG3} \\ [1ex]
$B^{\circ}\rightarrow K^{*}(892)^{\circ}\mu^{+}\mu^{-}$ & $2.3\times 10^{-5}$ &
	UA1 & \citenum{UA1}
\end{tabular}
\end{center}

\subsection{Model Dependency.}

Some of these limits, \eg,
$B\rarrow K^{\ast} \gamma$, are only
a factor of a few times that calculated. I will examine the
model dependency of the types of calculations that have been done so
far. We shall see that whereas the inclusive decays are very well
understood, the exclusive decays still have a lot of uncertainty.

The hadronic matrix elements relevant to the transition
\(B(b\bar{q})\rightarrow V(Q\bar{q})\), where $V$ is a vector meson,
are given by \cite{OT}
\begin{eqnarray}
\langle V(k)|\bar{Q}i\sigma_{\mu\nu}q^{\nu}b_{R}|B(k')\rangle &=&
f_{1}(q^{2})i\varepsilon_{\mu\nu\lambda\sigma}
\epsilon^{\ast\nu}k'^{\lambda}
k^{\sigma}\nonumber\\
&&+\left[(m^{2}_{B}-m^{2}_{V})\epsilon^{\ast}_{\mu}-
\epsilon^{\ast}\cdot q(k'+k)_{\mu}\right]f_{2}(q^{2})\nonumber\\
&&+\epsilon^{\ast}\cdot q\left[(k'-k)_{\mu}-\frac{q^{2}}{(m^{2}_{B}
-m^{2}_{V})}(k'+k)_{\mu}\right]f_{3}(q^{2})\!\,\, ,\\
\langle V(k)|\bar{Q}\gamma_{\mu}b_{L}|B(k')\rangle &=&
T_{1}(q^{2})i\varepsilon_{\mu\nu\lambda\sigma}\epsilon^{\ast\nu}
k'^{\lambda}k^{\sigma}
+(m^{2}_{B}-m^{2}_{V})T_{2}(q^{2})\epsilon^{\ast}_{\mu}
\nonumber\\
&&+T_{3}(q^{2})\epsilon^{\ast}\cdot q(k'+k)_{\mu}
+T_{4}(q^{2})\epsilon^{\ast}\cdot q(k'-k)_{\mu}\,\,\, .
\end{eqnarray}
where $q=k'-k$.

In the static $b$-quark limit \cite{IW1,BD}, the $b$-quark spinor has
only upper component in the $B$ rest frame. Thus in the $B$ rest frame,
we have the following relations between the $\gamma_{\mu}$ and
$\sigma_{\mu\nu}$ matrix elements:
\begin{eqnarray}
\langle V(k) | \bar{Q}\gamma_{i}b|B(k')\rangle &=&
\langle V(k) | \bar{Q}i\sigma_{0i}b|B(k')\rangle\,\,\, , \\
\langle V(k) | \bar{Q}\gamma_{i}\gamma_{5}b|B(k')\rangle &=&
- \langle V(k) | \bar{Q}i\sigma_{0i}\gamma_{5}b|B(k')\rangle\,\,\, ,
\end{eqnarray}
which relate the form factors $f_{1,2,3}$ to $T_{1,2,3,4}$ as
\begin{eqnarray}
f_{1}&=& -(m_{B}-E_{V})T_{1}-\frac{(m^{2}_{B}-m^{2}_{V})}{m_{B}}T_{2}
\,\,\, , \label{f1}\\
f_{2}&=& -\frac{1}{2}\left[ (m_{B}-E_{V})-(m_{B}+E_{V})
\frac{q^{2}}{m^{2}_{B}-m^{2}_{V}}\right]T_{1}
-\frac{1}{2m_{B}}\left( m^{2}_{B}-m^{2}_{V}+q^{2}\right) T_{2}
\!\,\, , \nonumber\\
&&\label{f2}\\
f_{3}&=& -\frac{1}{2}(m_{B}+E_{V})T_{1}+\frac{1}{2m_{B}}
(m^{2}_{B}-m^{2}_{V})(T_{1}+T_{2}+T_{3}-T_{4})\,\,\, . \label{f3}
\end{eqnarray}
where $E_{V}=(m^{2}_{B}+m^{2}_{V}-q^{2})/(2m_{B})$.

Now, we wish to consider the case where $V$ is the $K^{\ast}$.
The branching ratio of the exclusive process $B\rightarrow K^{\ast}\gamma$
to the inclusive process $b\rightarrow s\gamma$
can be written in terms of \(f_{1}\) and \(f_{2}\) at $q^{2}=0$, as
\cite{Alt,Desh}
\begin{equation}
R=\frac{\Gamma (B\rightarrow K^{\ast}\gamma)}
{\Gamma (b\rightarrow s\gamma)}
\cong \frac{m_{b}^{3}(m_{B}^{2}-m_{K^{\ast}}^{2})^{3}}
{m_{B}^{3}(m_{b}^{2}-m_{s}^{2})^{3}}\frac{1}{2}\left[|f_{1}(0)|^{2}
+4|f_{2}(0)|^{2}\right].
\label{ratio}
\end{equation}
Using the static $b$ quark limit, Eqs. (\ref{f1}) and (\ref{f2}), we
have $f_{2}(0)=(1/2)f_{1}(0)$. This is often referred to
as a quark model result but it is clear that it is more general than
that and requires only that the $b$ quark is heavy enough.

Although there is now only one form factor to calculate, this is
still a controversial calculation \cite{OT,Alt,Desh,Dom,Aliev} with there being
a factor of about ten uncertainty coming from the way in which the large
recoil of the $K^{\ast}$ is handled.

I will review the quark model calculations to show where the ambiguity arises.

The problem comes from the momentum wave functions.
These are determined by solving the Schroedinger equation of the corresponding
\(q\bar{q}\) system with a Coulomb plus linear potential\cite{isgw89,hayn82}
between the quarks. For $L$=0 meson states, which we will consider here, they
are chosen to be Gaussian wave functions of the form
\begin{equation}
\phi(\vec{p})=(\pi\beta^{2})^{-3/4}\,e^{-\vec{p}\,^{2}/2\beta^{2}}\,\, ,
\label{e3}
\end{equation}
\noindent
with a variational parameter $\beta$. The formulation of the
relative momentum wave function is then obviously non relativistic.
The \(q_{3}(\vec{k}+\vec{p})\bar{q}_{2}(-\vec{p})\) system of the $K^{\ast}$
becomes highly relativistic in the  region of large recoil
and the use of the above non relativistic
momentum wave function for \(\phi_{K^{\ast}}\) is then questionable.
In Ref. [\citenum{isgw89}] this problem was treated by
fixing the meson and quark spinor normalizations at the
zero recoil point and ignoring all of the recoil dependence in the
matrix element except for the momentum wave function part.
The recoil momentum can be written as
\begin{equation}
|\vec{k}|=\sqrt{E_{K^{\ast}}^{2}-M_{K^{\ast}}^{2}}=\sqrt{1+\frac{t_{m}-q^{2}}
{4M_{B}M_{K^{\ast}}}}\sqrt{\frac{M_{K^{\ast}}}{M_{B}}}\sqrt{t_{m}-q^{2}},\label{e4}
\end{equation}

\noindent
where \(q^{2}=(P_{B}-P_{K^{\ast}})^{2}\) and \(t_{m}\equiv (M_{B}
-M_{K^{\ast}})^{2}\). Since $q^{2}=t_{m}$ corresponds to the point
of zero recoil, we have
\(|\vec{k}|=\sqrt{M_{K^{\ast}}/M_{B}}\sqrt{t_{m}-q^{2}}\), the non
relativistic form for \(|\vec{k}|\) near this point.
This non relativistic form of the recoil
momentum in the momentum wave function was adopted in Ref. [\citenum{isgw89}]
and the recoil dependence of
the matrix element at large recoil was prescribed by multiplying
$|\vec{k}|$ with a relativistic correction factor $1/\kappa$
($\kappa=0.7$ was determined by fitting the
pion form factor to experiment\cite{beb78}).

Last year, we showed\cite{OT} that it is possible to do without
$\kappa$ if the recoil
dependence of the spinors is taken into account. A similar
result has now been given using Bjorken's sum rule in heavy
quark symmetry.  Nevertheless, the
problem remains that, depending on the choice of how the evaluation
of $|\vec{k}|$ is performed, there can be a factor of about five in
the result for the exclusive decay.

I have not been able to account for the extra factor of two for
the branching ratio in the sum rule approach\cite{Dom,Aliev}.

\subsection{Heavy Quark Symmetries.}

One way to overcome the model dependencies that plague the
exclusive calculations might be to try the heavy quark symmetry
approach\cite{IW1,VS,Eic,PW,IW2}. It has been pointed out\cite{IW1,IW2}
that heavy quark symmetries could relate the data on the semi-leptonic decays
\(D\rightarrow Ke^{+}\nu\) and \(D\rightarrow K^{\ast}e^{+}\nu\) and
provide information relevant to the
exclusive decay \(B\rightarrow K^{\ast}\gamma\).
However,  there is the
problem of continuation of these symmetries from zero recoil
momentum across the Dalitz plot to the largest recoil $q^{2}=0$.
It has been suggested\cite{isg90} that the relations among
the operator matrix elements might be valid over the full kinematic
ranges even for transitions of the type $b \rightarrow s$.

The question has been raised\cite{BJ} whether the $s$ quark is
sufficiently heavy to apply
these symmetries to the $K^{\ast}$. Here, I will show\cite{OT2} that the
important ingredient is that the $b$ quark is heavy and that
corrections are suppressed.

The heavy quark symmetries are derived in the large mass
limit. Many of the relations have been known to hold, at least in an
approximate sense, in the quark model. For example, in
the heavy quark model, the spin of $Q$ in $V(Q\bar{q})$ decouples from
the gluon, giving
\[
S^{Z}_{Q}|V(Q\bar{q})\rangle = \frac{1}{2}|P(Q\bar{q})\rangle
\,\,\, , \,\,\,
S^{Z}_{Q}|P(Q\bar{q})\rangle = \frac{1}{2}|V(Q\bar{q})\rangle
\,\,\, .
\]
where $S_{Q}$ is the spin operator of the $Q$ quark, and $P$ is
the scalar meson corresponding to $V$ with the same quark
content of $Q\bar{q}$. The matrix relations
\[
\langle V(k)|\bar{Q}\Gamma b|B(k')\rangle =
2 \langle P(k)| [S^{Z}_{Q},\bar{Q}\Gamma b]|B(k')\rangle \,\,\, ,
\]
for $\Gamma$ any product of $\gamma$-matrices, then gives
additional relations among the form factors
\begin{equation}
-T_{1}=2T_{3}=-2T_{4}
\end{equation}
These relations are also valid, to within a few percent, in the quark
model\cite{OT}.

Instead of seven, we now have two form factors,
$T_1$ and $T_2$, say. We further assume that the quarks are
sufficiently heavy so that in the equation of motion of the quarks we
can replace the quark masses with the meson masses,

\begin{equation}
\langle V(k)|\bar{Q}(k\!\!\! /-k'\!\!\!\! / ) \gamma_{5}b|B(k')\rangle
\approx -(m_{B}+m_{V})
\langle V(k)|\bar{Q}\gamma_{5}b|B(k')\rangle.
\end{equation}

Since $m_b\approx m_B$, and $m_B\geq m_V$, the error of using $m_V$ in place
of $m_Q$ is suppressed. The static $b$ limit,
$\langle V(k)|\bar{Q} \gamma_{5}b|B(k')\rangle =
-\langle V(k)|\bar{Q}\gamma_{0} \gamma_{5}b|B(k')\rangle$,
lets us relate $T_2$ to $T_1$;

\begin{equation}
(m_{B}^{2}-m_{V}^{2})\frac{T_2}{T_1}=\frac{1}{2}\left[ (m_B+m_V)^{2}-q^{2}
\right]
\end{equation}

The symmetries relate the form factors in the
following ways\cite{OT}:
\begin{eqnarray}
T_{1}&=&\frac{-2(m_{B}^{2}-m_{K^{\ast}}^{2})}
{(m_{B}+m_{K^{\ast}})^{2}-q^{2}}\,T_{2}
=\, 2T_{3}
\rule{0cm}{1.4cm}
=\, -2T_{4}\nonumber\\
&=&\frac{-1}{(m_{B}+m_{K^{\ast}})}\,f_{1}
=\frac{2(m_{B}+m_{K^{\ast}})}{(m_{B}+m_{K^{\ast}})^{2}-q^{2}}\,f_{2}
=\frac{2}{m_{B}-m_{K^{\ast}}}\,f_{3}\nonumber\\ \label{e20}
\end{eqnarray}
The presence of masses here show the effect of mass breaking of the
heavy quark symmetries. On the other hand, the above discussion
explicitly shows that the errors are controlled, even for a $K^{\ast}$, which
normally would not be thought to be a good candidate for the heavy quark
theory.
A plot\cite{OT} of individual terms in Eq. (\ref{e20}) (times a
normalization factor $\sqrt{4m_{B}m_{K^{\ast}}}$) shows that the equalities
hold very well, with less than about a 15\% discrepancy, across
the whole kinematic region to $q^{2}=0$.
The small discrepancy reflects the error of using $m_V$ in place
of $m_Q$, as noted above.

However, there still remains the troublesome
problem of what to do with the overlap function. A start to solving
this was proposed by Burdman and Donoghue\cite{BD} in which the heavy
quark symmetry arguments were used to relate
$B\rightarrow K^{\ast}\gamma$ to the semileptonic process
$B\rightarrow \rho e \bar{\nu}$ using the static $b$-quark
limit and $SU(3)$ flavor symmetry.
The problem with this is that the semileptonic decay
vanishes at the kinematic point they use.
This means that experimentally there should be no event at that point
and very few in the neighbourhood, causing a large uncertainty in
the measurement.
Recently a new relation between the branching ratio
$R(B\rightarrow K^{\ast}\gamma)$ and the $q^{2}$-spectrum for
$B\rightarrow\rho e\bar{\nu}$ has been given\cite{OT2}.
A direct measurement of
$d\Gamma (B\rightarrow\rho e\bar{\nu})/dq^{2}$ at $q^{2}=0$ can
therefore provide relevant information for
$R(B\rightarrow K^{\ast}\gamma)$ since the $q^{2}$-spectrum
for $B\rightarrow\rho e\bar{\nu}$ does not vanish at $q^{2}=0$.

\bibliographystyle{unsrt}

\end{document}